\documentclass[%
 reprint,
superscriptaddress,
 amsmath,amssymb,
 aps,
]{revtex4-2}

\usepackage{hyperref}
\usepackage{braket}
\usepackage{graphicx}
\usepackage{dcolumn}
\usepackage{bm}

\def\GG{{\overleftrightarrow{{\bf G}}}}

\begin{document}

\preprint{APS/123-QED}

\title{Exploiting oriented field projectors to open topological gaps \\ in plasmonic nanoparticle arrays}

\author{Álvaro Buendía}
\email{a.buendia@csic.es}
\affiliation{Instituto de Estructura de la Materia, Consejo Superior de Investigaciones Científicas, Serrano 121, 28006 Madrid, Spain}

\author{José A. Sánchez-Gil}%
\affiliation{Instituto de Estructura de la Materia, Consejo Superior de Investigaciones Científicas, Serrano 121, 28006 Madrid, Spain}

\author{Vincenzo Giannini}
\affiliation{Instituto de Estructura de la Materia, Consejo Superior de Investigaciones Científicas, Serrano 121, 28006 Madrid, Spain}
\affiliation{Technology Innovation Institute, Masdar City 9639, Abu Dhabi, United Arab Emirates}
\affiliation{Centre of Excellence ENSEMBLE3 sp. z o.o., Wolczynska 133, Warsaw, 01-919, Poland}	

\begin{abstract}
In the last years there have been multiple proposals in nanophotonics to mimic topological condensed matter systems. However, nanoparticles have degrees of freedom that atoms lack of, like dimensions or shape, which can be exploited to explore topology beyond electronics. Elongated nanoparticles can act like projectors of the electric field in the direction of the major axis. Then, by orienting them in an array the coupling between them can be tuned, allowing to open a gap in an otherwise gapless system. As a proof of the potential of the use of orientation of nanoparticles for topology, we study 1D chains of prolate spheroidal silver nanoparticles. We show that in these arrays spatial modulation of the polarization allows to open gaps, engineer hidden crystalline symmetries and to switch on/off or left/right edge states depending on the polarization of the incident electric field. This opens a path towards exploiting features of nanoparticles for topology to go beyond analogues of condensed matter systems.
\end{abstract}
\maketitle
\section{Introduction} 
The exciting discovery of the topological phase of matter systems has inspired many new fields in physics, particularly in photonics~\cite{Lu2014,Khanikaev2017,Yves2017} ; in fact, in recent years, we have witnessed an exponential growth of interests in that direction. Mimicking the phenomenology of topological insulators has been the driving force until now. However, it is becoming clear that a further step needs to be taken, i.e., to push forward new topological photonic phenomenology that does not have a material counterpart.

Topological insulators are possible thanks to the fermionic nature of electrons \cite{Kane2005}, but photons cannot take advantage of such symmetry. Initial solutions have been proposed based on gyromagnetic  photonic crystals \cite{Wang2008}, 
bianisotropic materials  \cite{Khanikaev2012}, and 
coupled waveguides and resonators \cite{Hafezi2011,Rechtsman2013}.

All the previous systems use some kind of time-reversal property not present in simple photonics systems without magnetic response. In addition, there is always a strong interest in achieving very small and faster devices for nanotechnological applications. Typical examples are microprocessors, but light interacts weakly with the material at the nanoscale. Moreover, one would like to have such photonic properties in the visible, where most of the molecular electronic transitions happens, making such zone relevant in light-matter interaction. With these goals and restriction in mind, metal nanoparticles using plasmonic resonances are probably the best candidates.
This has made it possible for many researchers to look at what we can call topological nanoparticle photonics \cite{Rider2019, Rider2022}.

Plasmonic nanoparticles provide an excellent platform for light-matter interaction, but being not simple to break time-reversal symmetry in the visible range, a typical approach is using crystal symmetries~\cite{Wu2015, Siroki2017, Peng2019, Parappurath2020, Liu2020, Palmer2021}. Such approach has also been explored for radiative heat transfer with interesting results~\cite{Sanders2021}.

In addition, particular care needs to be taken due to the long-range nature of these interactions and the radiative corrections \cite{Pocock2018,Pocock2019, Rider2022}, which can spoil the topological protection of the system. Here, instead of focusing on such loss of protection, we explore degrees of freedom of the nanoparticles which could be exploited for topology beyond condensed matter systems. 

This paper is organized as follows. In section \ref{section:SSHmodel} we introduce the simplest topological system, a dimer chain known as SSH model, and its extensions for larger unit cells. In section \ref{section:Topoplasm} we study and compare different plasmonic counterparts of these chains. Finally, in section \ref{section:Excitation} we discuss the excitation and switching of the edge states of the plasmonic chains of elongated nanoparticles by an incoming electric field.

\section{SSH model and topological phases}
\label{section:SSHmodel}

\begin{figure}[t]
\begin{center}
\includegraphics[width=0.5\textwidth]{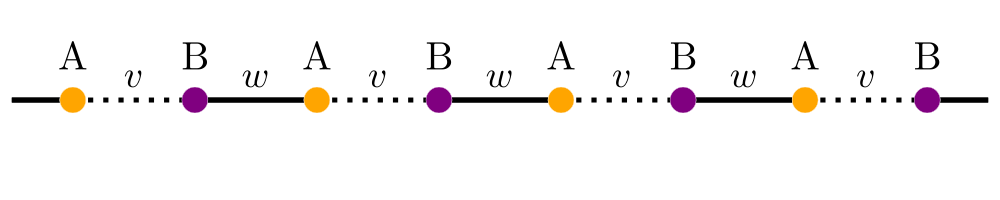}
\includegraphics[width=0.5\textwidth]{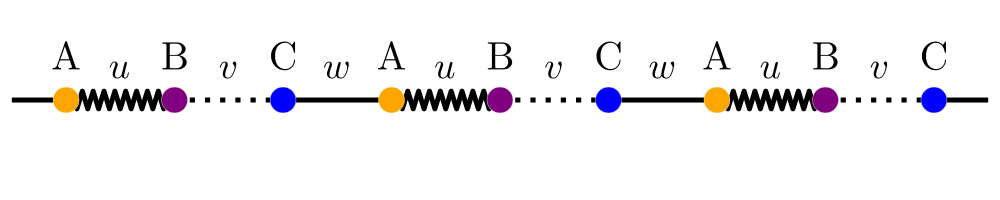}
\includegraphics[width=0.5\textwidth]{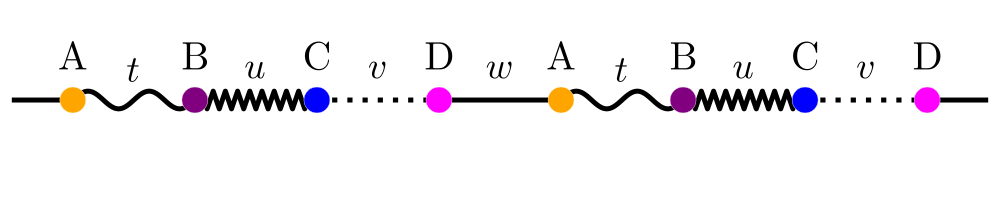}
\end{center}
\caption{(a) The simplest model with topological protection, the SSH chain, consisting of a diatomic chain with staggered hoppings $v$ and $w$. (b) Extended 3-particle SSH chain (or SSH3), alternating hoppings $u,v$ and $w$. (c) Extended 4-particle SSH chain (or SSH4), alternating hoppings $t,u,v$ and $w$.}
\label{fig:tbchains}
\end{figure}

The Su-Schrieffer-Heeger (SSH) model is the simplest system with topological protection. It was firstly proposed in Ref. \cite{Su1979} to describe the physics of the polyacetylene chain, which alternates double and simple (or strong and weak) bonds between adjacent carbon atoms. In FIG.~\ref{fig:tbchains}(a) we show a scheme of this model, which consists of a one-dimensional diatomic chain with two staggered hopping amplitudes between nearest neighbors, namely $v$ and $w$. Its tight-binding Hamiltonian is: 

\begin{align} 
\mathcal{H}(q) =  \begin{pmatrix} 0 & v+w e^{-iqd} \\ v+w e^{iqd} & 0 \end{pmatrix},
\label{eq:BlochSSH}
\end{align}

and satisfies the Schrödinger equation:
\begin{align} 
\mathcal{H}(q)\ket{u_n(q)} = E_n(q) \ket{u_n(q)}
\label{eq:SchrödingerBloch}
\end{align}

This system has two different distinct topological phases: it is trivial when the bond between particles in adjacent unit cells is weaker than the one between particles within a unit cell ($|w|<|v|$) and topological when it is stronger ($|w|>|v|$). When we chop the periodic chain commensurately with the topological unit cell, it hosts strongly localized zero-energy states at both ends. These edge states are robust to disorder and perturbations that respect its symmetries: sublattice symmetry (also known as chirality) and mirror/inversion symmetries.
Sublattice symmetry stems from the existence of two sublattices ($A$ and $B$) with bonds between sublattices but not within a sublattice. This implies the Hamiltonian is anti-block-diagonal, i.e.:
\begin{equation}
\mathcal{H}(q) = \begin{pmatrix} 0 & h(q) \\ h^\dag(q) & 0  \end{pmatrix}.
\end{equation}
Chirality makes the spectrum symmetric around zero energy and fixes the energy of the edge states  at zero, isolating them from the bulk states. It also makes each edge state be localized in just one of the sublattices.
\begin{equation}
\Gamma \mathcal{H}(q) \Gamma = -\mathcal{H}(q) \qquad \Gamma = \begin{pmatrix} I & 0 \\ 0 &-I \end{pmatrix}
\end{equation}
The SSH model is mirror-symmetric because the system remains invariant under spatial inversion in the $x$-axis, i.e., under the operation of $\Pi$:
\begin{equation}
\Pi \mathcal{H}(q) \Pi = \mathcal{H}^T(-q) = \mathcal{H}(q) 
\end{equation}

The SSH chain is also inversion-symmetric, as it remains invariant under the subsequent spatial inversion in all axes. Mirror or inversion symmetries lead to the double degeneracy of the edge states, even when the sublattice symmetry is broken.  

As long as chirality and/or mirror symmetries are respected, the Zak phase \cite{Zak1989} $\gamma$, a bulk property, is a topological invariant and predicts the existence of edge states in the terminated system \cite{Cao2017}. This is known as bulk-boundary correspondence.
The Zak phase for each band and for the gap is:
\begin{align}
    \gamma_n &= \int_{\mathrm{BZ}} \langle u_n (q) | \frac{\partial}{\partial_q} | u_n (q) \rangle dq \nonumber \\
\gamma &= \sum_{\textnormal{below gap}} \gamma_n
\label{eq:Zak}
\end{align}
In electronic systems, when the system is neutral, only half of the bands are below the Fermi level, so in the SSH model only the lower band contributes to the Zak phase. 

In the following subsection we introduce some extensions of the SSH model. 

\subsection{Extended unit cell SSH models}

Due to the simplicity of the system, several generalizations of the SSH model have been made, for example by adding hoppings between further neighbors \cite{PerezGonzalez19}, or by extending to two dimensions in a square array \cite{Kim2020}. This model can also be generalized to one-dimensional chains with larger linear unit cells \cite{Maffei2018,Bid2022,Zhang2021,MartinezAlvarez19} or rhombus unit cells \cite{Li2021}. These systems are topologically more complex than the SSH model, featuring several gaps and non-zero edge states. They  can also exhibit other kinds of topological protection like square-root topology \cite{Arkinstall17}. 

First, as we show in FIG.~\ref{fig:tbchains}(b), we consider a linear chain with three alternating hopping amplitudes $u,v$ and $w$ This lattice has three different sublattices $A,B$ and $C$. The topology of this system has been discussed in Refs. \cite{Anastasiadis22, MartinezAlvarez19, Liu2017}
\begin{align} 
\mathcal{H}(q) =  \begin{pmatrix} 0 & u & w e^{-iqd} \\ u & 0 & v \\ w e^{iqd} & v & 0 \end{pmatrix}.
\label{eq:BlochSSH3}
\end{align}
A three-way  generalization of the sublattice symmetry can be made for this chain: 
\begin{eqnarray} 
 \Gamma_3^{-1}\mathcal{H}(q)\Gamma_3+ \Gamma_3^{-2}\mathcal{H}(q)\Gamma_3^2 = -\mathcal{H}(q), \nonumber \\
\Gamma_3 = \begin{pmatrix} 1 & 0 & 0 \\ 0 & e^{2i\pi/3} & 0 \\ 0 & 0 & e^{-2i\pi/3} \end{pmatrix} , \qquad \Gamma_3^3 = I.
\end{eqnarray}
This symmetry is the same that features the breathing Kagome lattice \cite{Herrera2022}. However, due to the absence of the $C_3$ rotational symmetry also present in the Kagome lattice, in this 1D system there are no three degenerate zero-states but non-zero edge states. Additionally, the edge states are not localized in just one of the sublattices but the two closer to the edge. When the chain is mirror symmetric, i.e $|u|=|v|$, the edge states come in two degenerate pairs at energies $-E$ and $E$ when $|w|>|u|$. 
Each gap has a distinct Zak phase, that is quantized by mirror symmetry. Due to the three-way chirality they are not independent, but the same.  

Similarly, we can consider a chain with a 4-particle unit cell, shown in FIG.~\ref{fig:tbchains}(c), with hoppings $t,u,v$ and $w$. The tight-binding Hamiltonian (in the base $A,C,B,D$) is: 
\begin{align} 
\mathcal{H}(q) =  \begin{pmatrix} 0 & 0 & t  & w e^{-iqd} \\ 0& 0 & u & v \\ t & u & 0 & 0\\ w e^{iqd} & v & 0 & 0 \end{pmatrix}
\label{eq:BlochSSH4}
\end{align}
Due to the even number of particles in the unit cell, this chain recovers the sublattice symmetry from the SSH model.
The transition for the central gap occurs when $|tv|=|uw|$. When $|tv|>|uw|$ the system is in the trivial phase, whereas for $|tv|<|uw|$ the system has symmetry-protected zero-energy states that localize exclusively in even or odd sublattices. 

However, the system also features non-zero energy states in the lower/upper gaps. These states inherit properties from the four-way generalized chirality, which is given by:
\begin{eqnarray} 
 \Gamma_4^{-1}\mathcal{H}(q)\Gamma_4+ \Gamma_4^{-2}\mathcal{H}(q)\Gamma_4^2 +\Gamma_4^{-3}\mathcal{H}(q)\Gamma_4^3 = -\mathcal{H}(q),  \nonumber \\
\Gamma_4 = \begin{pmatrix} 1 & 0 & 0 & 0  \\ 0 & i & 0& 0 \\ 0 & 0 & -1 & 0 \\ 0 & 0 & 0 &-i \end{pmatrix}, \qquad \Gamma_4^4 = I.
\end{eqnarray}
This symmetry makes the non-zero edge states be localized in three out of the four sublattices. However, this symmetry doesn't close lower/upper gaps or quantize their Zak phases, but spatial symmetries do. The system is mirror/inversion symmetric when $|t| = |v|$. Mirror symmetric non-zero energy edge states appear for $|w|>|u|$.

Now we focus in arrays of plasmonic nanoparticles. Previously, optical response of metallic nanoparticles has been used to mimic topological condensed matter systems, as in zigzag chains \cite{Poddubny2014, Zhang21(2)}, diatomic chains of nanospheres \cite{Downing2017}, or breathing Kagome \cite{Proctor2021} and breathing honeycomb plasmonic metasurfaces \cite{Honari2019, Proctor2020}. However, in these systems it has been shown that long-range  interactions between nanoparticles must be considered, which have a striking effect on the topology of the system~\cite{Pocock2018, Pocock2019}.

Electric fields produce localized surface plasmon resonances (LSPR) in metallic nanoparticles. A small single nanoparticle with $a \ll \lambda$ (where $a$ is the particle radius and $\lambda$ is the wavelength of incoming light), such that $a> 3-4$ nm to avoid quantum effects, scatters an incident electric field $\textbf{E}_{inc}$ approximately like a dipole $\textbf{p}$: 
\begin{align}
    \mathbf{p} = \epsilon_{\mathrm{B}}\overleftrightarrow{\alpha}(\omega) \mathbf{E}_{inc}, 
    \label{eq:dipole_Efield}
\end{align}
where $\epsilon_B$ is the permittivity of the background medium and $\overleftrightarrow{\alpha}(\omega)$ is the polarizability tensor.

The dipolar approximation still holds for an array of nanoparticles, if they are separated a distance of at least $3a$. 

Then, each dipole in the array is determined by  both the incident electric field and the scattered electric field by the rest of the  dipoles: 
\begin{align}
   \mathbf{p}_n &= \overleftrightarrow{\alpha_n}(\omega)\left(\textbf{E}_{inc} + \frac{k^2}{\epsilon_0}\sum_{m\neq n}  \overleftrightarrow {\mathbf{G}}(\mathbf{r}_n,\mathbf{r}_m,\omega)\mathbf{p}_m\right),
\label{eq:dipoles}
\end{align}

where $n,m$ are sites in the array, $\mathbf{p}_{n,m}$ are the dipoles in positions $\mathbf{r}_{n,m}$ and $\overleftrightarrow{\mathbf{G}}(\mathbf{r}_n,\mathbf{r}_m,\omega)$ is the Green's dyadic function, which in the quasi-static regime $kR\gg 1$ is given by:
\begin{align}
\GG(\mathbf{r}_n,\mathbf{r}_m,\omega) &= \frac{1}{4\pi k^2 R^3}\left[-I+3\frac{\mathbf{R}\otimes \mathbf{R}}{R^2}\right],
\label{eq:dyadic_G2}
\end{align}
where  $\mathbf{R}=\mathbf{r}_n-\mathbf{r}_m$, $R=|\mathbf{R}|$, and $k=\sqrt{\epsilon_B} \omega/c$ is the wavevector. 

In the next subsections we'll consider examples of topological arrays of nanoparticles.

\section{Topological plasmonic chains}
\label{section:Topoplasm}

\subsection{Chain of nanospheres} 

First, we consider a single spherical metallic nanoparticle. A nanosphere has spherical symmetry, so its tensor polarizability behaves like a scalar, $\alpha(\omega)$, which in the quasi-static limit $a\ll \lambda$ is: 

\begin{align}
\alpha(\omega) = 4\pi a^3 \epsilon_0 \frac{\epsilon(\omega) - \epsilon_\mathrm{B}}{\epsilon(\omega) + 2\epsilon_\mathrm{B}}. 
\label{eq:alpha_QS}\end{align}

In figure \ref{fig:nanoparticles}(a) we see the optical response of a single silver nanosphere of radius $a=12.5$ nm to a linear-polarized electric field (blue curve), that shows a resonance for $\hbar\omega_{sp}\sim 2.75$ eV. 

Now we consider a plasmonic analogue of the SSH model, i.e., a chain of nanospheres with two alternate distances: the intracell distance $\beta\frac{d}{2}$ and the intercell distance $\left(2-\beta\right)\frac{d}{2}$, $d$ being the size of the unit cell (see scheme on FIG.~\ref{fig:nanoparticles}(b)).
For any array of nanospheres and in absence of incident electric field we can rewrite EQ.~\ref{eq:dipoles} as:
\begin{equation}
\frac{1}{\alpha(\omega)} \textbf{p}_n =  \frac{k^2}{\epsilon_0} \sum_{m \neq n} \overleftrightarrow{\textbf{G}}(\textbf{r}_n,\textbf{r}_m, \omega)\cdot \textbf{p}_m.
\label{eq:eqGiso}
\end{equation}

After Bloch, coupled-dipole equations can be compacted in a matrix equation:
\begin{equation}
\mathcal{G}(\textbf{q}) \textbf{P} = \frac{1}{\alpha(\omega)} \textbf{P},
\label{eq:CDBloch}
\end{equation}
where $\textbf{P} = (p_{1x},p_{2x},p_{1y},p_{2y}, p_{1z}, p_{2z})$~\cite{Pocock2018}. As we see, this is equivalent to the Schrödinger equation in EQ. \ref{eq:SchrödingerBloch}, where the dipole vector, the inverse of the polarizability, and the Green's matrix $\mathcal{G}(q)$ take the roles of, respectively, eigenvectors, eigenvalues, and the Bloch Hamiltonian $\mathcal{H}(q)$. Explicitly, $\mathcal{G}(q)$ terms are: 
 \begin{figure}[h]
\begin{center}
\includegraphics[width=0.5\textwidth]{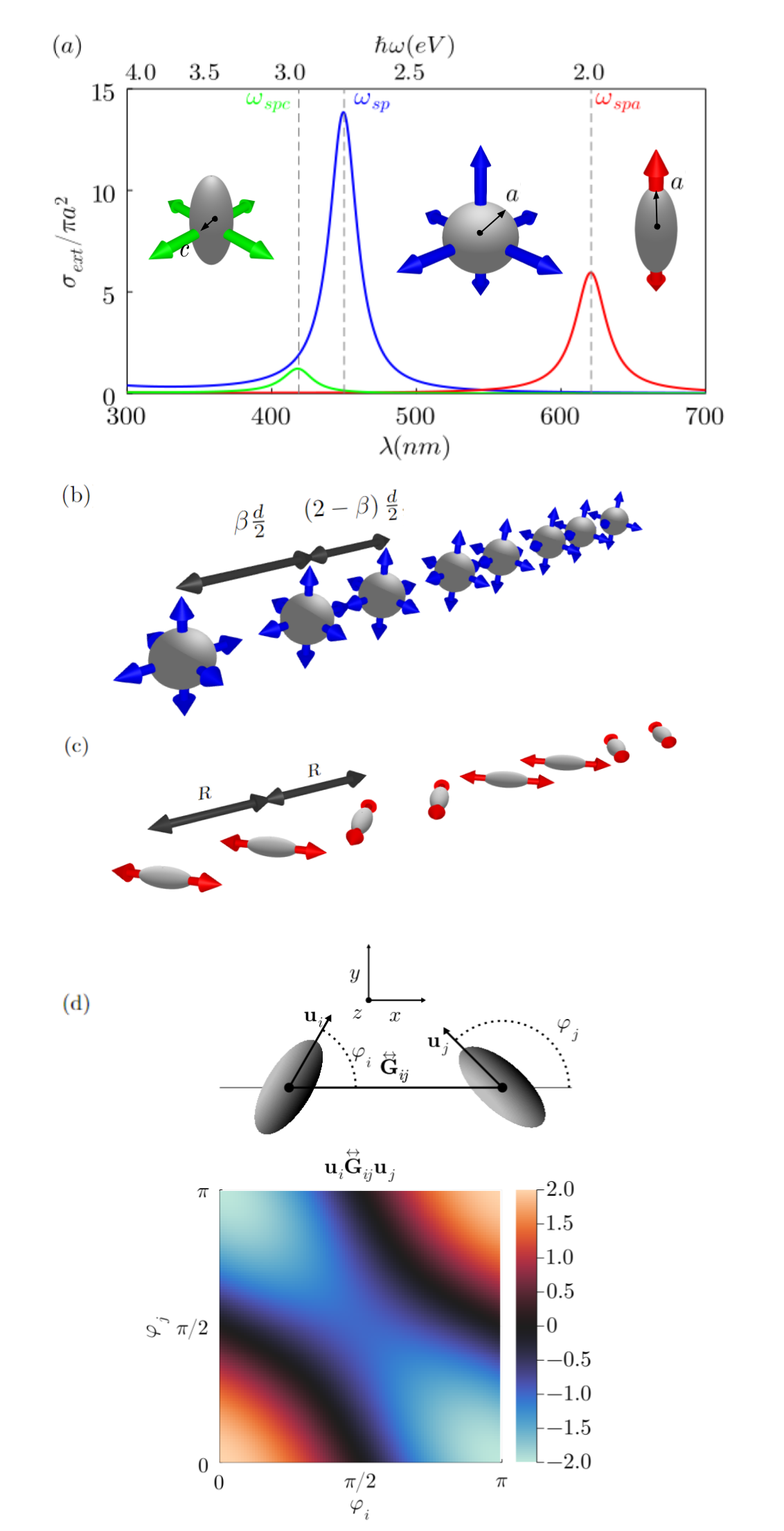}
\end{center}
\caption{\textbf{Arrays of plasmonic nanoparticles} (a) Extinction cross sections of a silver nanosphere of radius $a=12.5$ nm (blue curve) and prolate spheroidal ($a=12.5$ nm, $b=c=5$ nm) silver nanoparticles (red curve for the major axis polarization and green line for any polarization in the perpendicular plane) embedded in glass ($\epsilon_B = 2.25$); Ag dielectric function from \cite{Rodrigo2008}. (b-c) Schematic of the SSH analogue consisting of plasmonic chains with (b) alternating distances between nanospheres  and  (c) equidistant nanospheroids alternating orientations. (d) Interaction between nanospheroids depending on their orientations. Projection of the Green dyadic's function on the directions of the nanoparticles normalized by $\frac{1}{4\pi k^2 R^3}$.}
\label{fig:nanoparticles}
\end{figure}
\begin{align} 
\mathcal{G}_{\mu\nu} = \delta_{\mu\nu} \frac{2m_\nu}{\pi d^3} \begin{pmatrix} 0 & \frac{1}{\beta^3} + \frac{1}{(2-\beta)^3} e^{-iqd} \\ \frac{1}{\beta^3} + \frac{1}{(2-\beta)^3} e^{iqd} & 0 \end{pmatrix},
\label{eq:Gmunu}
\end{align}

\subsection{Coupled-dipole equations}

where $\mu,\nu = x,y,z$ are the polarizations of each pair of dipoles, and $m_v$ is $m_x = 2$ for the longitudinal polarization and $m_{y,z} = -1$ for the transversal polarizations. This is, the plasmonic dimer chain of nanospheres is equivalent to three independent copies of the SSH (EQ.~ \eqref{eq:BlochSSH}), one per polarization, with $v=\frac{2m_\nu}{\pi\epsilon_0 \beta^3 d^3}$ and $w=\frac{2m_\nu}{\pi\epsilon_0 (2-\beta)^3 d^3}$.
The dispersion bands $\omega(q)$ can be calculated from EQ.~\ref{eq:CDBloch}, searching for the solutions of: 
\begin{equation}
\lambda_n-\frac{1}{\alpha(\omega)} = 0.
\end{equation}
$\lambda_n$ being the $n$th eigenvalue of $\mathcal{G}(q)$.
The zero-energy modes typical of the finite SSH model translate in this system to six (two per polarization) resonant modes localized at the edges of the chain at the surface plasmon frequency of a single nanosphere $\omega_{sp}$.

Apart from the plasmonic diatomic 1D chain, the zigzag chain \cite{Poddubny2014,Zhang21(2)}, which alternates angles between the nanoparticles, has also been proposed to mimic the topology of the SSH model. This system exploits the polarization asymmetry between longitudinal and transversal modes and allows to select edge modes by changing the polarization of the incoming electric field. In the next subsection we will get advantage of this same anisotropy not in the geometry of the array but in the shape of the nanoparticles, adding a new degree of freedom to the system.

\subsection{Chain of nanospheroids} 

When the nanoparticles are not spherical, $\overleftrightarrow{\alpha_i}(\omega)$ are not proportional to the identity matrix anymore, so they affect the polarization of the dipoles. This asymmetry has no analogy in tight binding models and can be exploited to explore new topological systems. 
For example, if we replace the nanospheres in the previous chain by parallel nanorods, we can filter modes by in-plane or out-of-plane polarizations. However, by orienting the nanoparticles in alternating directions, we can force modes beyond the plasmonic nanosphere chain.  

Previously, gradual change of orientation in arrays of anisotropic nanoparticles or nanoholes has been exploited in Pancharatnam-Berry metasurfaces (also known as geometric phase metasurfaces) for example to enable polarization-dependent control of light  \cite{Huang2012} or to create vortex beams \cite{Yue2016}. 

Here we use orientation of elongated nanoparticles not to build a geometric phase in polarization, but to tune the interactions between nanoparticles in order to open a topological gap. However, the spatial modulation of polarization in our system plays a role in the control of edge states, allowing to switch them off by changing the polarization of the incident electric field, as we'll see in last section.

The strategy pursued in this paper to open a topological gap is similar to in Ref.~\cite{He2022}, where it is opened by orientation in the transversal plane of bianisotropic particles in an equidistant array. However, our study is more general and in the visible range. 

Let us now assume our elongated nanoparticles are prolate spheroidal nanoparticles with axis half-lengths $a >b = c$ with major axis pointing in the z direction. The results will be qualitatively equivalent for any other elongated shape, so we're not losing generality by making this choice. In this case the polarizability tensor is: 
\begin{equation}
\overleftrightarrow{\alpha}(\omega) = \begin{pmatrix} \alpha_{c}(\omega)  & 0 & 0  \\ 0 & \alpha_{c}(\omega) & 0 \\ 0 & 0 & \alpha_{a}(\omega) \end{pmatrix} ,
\end{equation}
where the quasistatic polarizabilities $\alpha_l$ with $l \in [a,b,c]$ are \cite{Moroz2009}:
\begin{equation}
\alpha_{l}(\omega) = V \frac{\epsilon(\omega)-\epsilon_b}{\epsilon_b + L_l(\epsilon(\omega)-\epsilon_b)} ,
\end{equation}
$V$ being  the volume of the spheroid, $V = \frac{4}{3}\pi a c^2$,and $L_l$ are geometric factors (see appendix \ref{section:AppxA}).

In FIG.~\ref{fig:nanoparticles}(a) we can see the extinction cross section (see \ref{section:AppxA} for details) of a single silver nanospheroid with major axis $a=12.5$ nm and minor axes $b=c=0.4a=5$ nm. Red curve represents the response to a field polarized parallel to the major axis, while for the green curve the field is polarized normal to the major axis. As we see, the resonance wavelengths are separate enough ($\hbar\omega_{spc} \sim 2.96 $ eV, whereas $\hbar \omega_{spa} \sim 2.0 $ eV), that in the proximity of the major axis resonance,  $\alpha_c(\omega \simeq \omega_{spa})\simeq 0$, so we can approximate the tensor polarizability to:
\begin{equation}
\overleftrightarrow{\alpha}(\omega \simeq \omega_{spa}) \simeq \begin{pmatrix} 0 & 0 & 0  \\ 0 & 0 & 0 \\ 0 & 0 & \alpha_{a}(\omega) \end{pmatrix}.
\end{equation}
    This means the polarizability tensor acts like a projection operator \cite{Kuntman18}, projecting the dipole in the direction of the major axis.  

Now  we consider an array of nanospheroids, with major axis oriented in the directions $\textbf{u}_n= (\cos{\varphi_n}, \sin{\varphi_n}, 0)^T$, where $\varphi_n$ is the angle of the particle $n$ with respect to the $x$ axis. We assume all the particles are oriented in the plane $xy$ for the sake of simplicity.  As the $y$ and $z$ axes are indistinguishable, all results will be the same for the $xz$ plane.  The equations for the general case, with nanoparticles oriented in any direction, are in appendices \ref{section:AppxA},\ref{section:AppxB}.

Due to the projection of the polarizability in the directions of the major axes of the particles, the vectorial coupled-dipole equations in an array of nanospheroids can be reduced to scalar equations: 
\begin{equation}
\frac{1}{\alpha_a(\omega)} p_n = \frac{k^2}{\epsilon_0}\sum _{m\neq n} G_{\textbf{u}_m, \textbf{u}_n} p_m ,
\label{eq:eqG2}
\end{equation}
where $G_{\textbf{u}_m, \textbf{u}_n} = (\GG(\textbf{r}_{m},\textbf{r}_{n},\omega)\cdot  \textbf{u}_m) \cdot \textbf{u}_n $ is the projection of the polarizability tensor in the directions $\textbf{u}_m$,$\textbf{u}_n$.

The coupled-dipole equations can again be rewritten in a matrix form:
\begin{equation}
\mathcal{G}_a(\textbf{k}) \textbf{P}_a = \frac{1}{\alpha_a(\omega)} \textbf{P}_a,
\end{equation}
where in this case $\mathcal{G}_a(\textbf{k})$ is a $N \times N$ matrix with elements given by $G_{\textbf{u}_m,\textbf{u}_n}$ and $\textbf{P}_a = (p_{1},...,p_{N})$ is a vector containing the module of the dipoles.   

First, let us consider a 1D chain of equidistant nanoparticles, separated by a distance $R$. We can see an scheme of this chain in FIG.~\ref{fig:nanoparticles}(c).  As all the particles are in the $x$ axis, the Green dyadic's function reduces to:
\begin{equation}
\overleftrightarrow{\textbf{G}}(\textbf{r}_m,\textbf{r}_n) = \frac{1}{4\pi k^2 R^3} \begin{pmatrix} 2 & 0 & 0 \\ 0 & -1 & 0 \\ 0 & 0 & -1 \end{pmatrix},
\label{eq:Gdyadic1d}
\end{equation} 
and its projection onto the $\textbf{u}_i$ and $\textbf{u}_j$ axes is:
\begin{equation}
G_{\textbf{u}_m, \textbf{u}_n} = \frac{
2\cos{\varphi_m}\cos{\varphi_n} - \sin{\varphi_m}\sin{\varphi_n}
}{4\pi k^2 R^3}.
\label{eq:eqGth1d}
\end{equation}

{\emph{This means that we can tune the coupling between nanoparticles by rotating them.}} 
In FIG.~\ref{fig:nanoparticles}(d) we plot the interaction $G_{\textbf{u}_m, \textbf{u}_n}$ multiplied by $4\pi k^2 R^3$ depending on the orientation angles $\varphi_m$ and $\varphi_n$. This interaction ranges from $2$ (parallel dipoles in the longitudinal direction) to $-2$ (antiparallel and in the longitudinal direction), passing by $-1$ (parallel dipoles in the transversal polarization). For any pair of angles in the zero contour line the interaction is suppressed. For example, orthogonal nanospheroids oriented in $x$ and $y$ directions don't interact between them, as we see for $\varphi = 0$ in the colormap. 

This zero interaction was impossible to achieve in the nanosphere chain and it could only be approximated by separating the particles a long distance (see EQ.~\ref{eq:Gmunu}). This could be interesting for switching off some even neighbor interactions that can break sublattice symmetry, but here we will restrict to the first-neighbor approximation. This approximation is accurate only in the quasi-static regime, that is, when the nanoparticles and the distances between them are small, so $kR\ll 1$.

In the next subsections we will consider linear arrays of nanospheroids. With nanospheres, a linear chain where the nanoparticles were equidistant would be gapless and equivalent to the $v=w$ case in the SSH model. However, by substituting the nanospheres by nanospheroids and adding the orientation as a degree of freedom  a gap can be opened. 

\subsubsection{2 particles/unit cell}
First we consider the simplest system which may have a topological gap. This is a linear array with two particles/unit cell and with a distance between adjacent particles $R=\frac{d}{2}$, being $d$ the size of the unit cell.  
For nanospheroids, the Green matrix of the system would be:
\begin{equation}
\mathcal{G}(q) = G_{\textbf{u}_A, \textbf{u}_B} \begin{pmatrix}0 & 1 + e^{-iqd} \\  1 + e^{iqd} & 0  \end{pmatrix}.
\end{equation}
This system is equivalent to the SSH model with $v = w = G_{u_A,u_B}$, due to the reciprocity $G_{u_A,u_B} = G_{u_B,u_A}$. This means that by rotating the spheroidal nanoparticles in a 2-particle unit cell, we can change the amplitude of the bands but not open a gap. 

\subsubsection{3 particles/unit cell}

\begin{figure}[h]
\begin{center}
\includegraphics[width=0.5\textwidth]{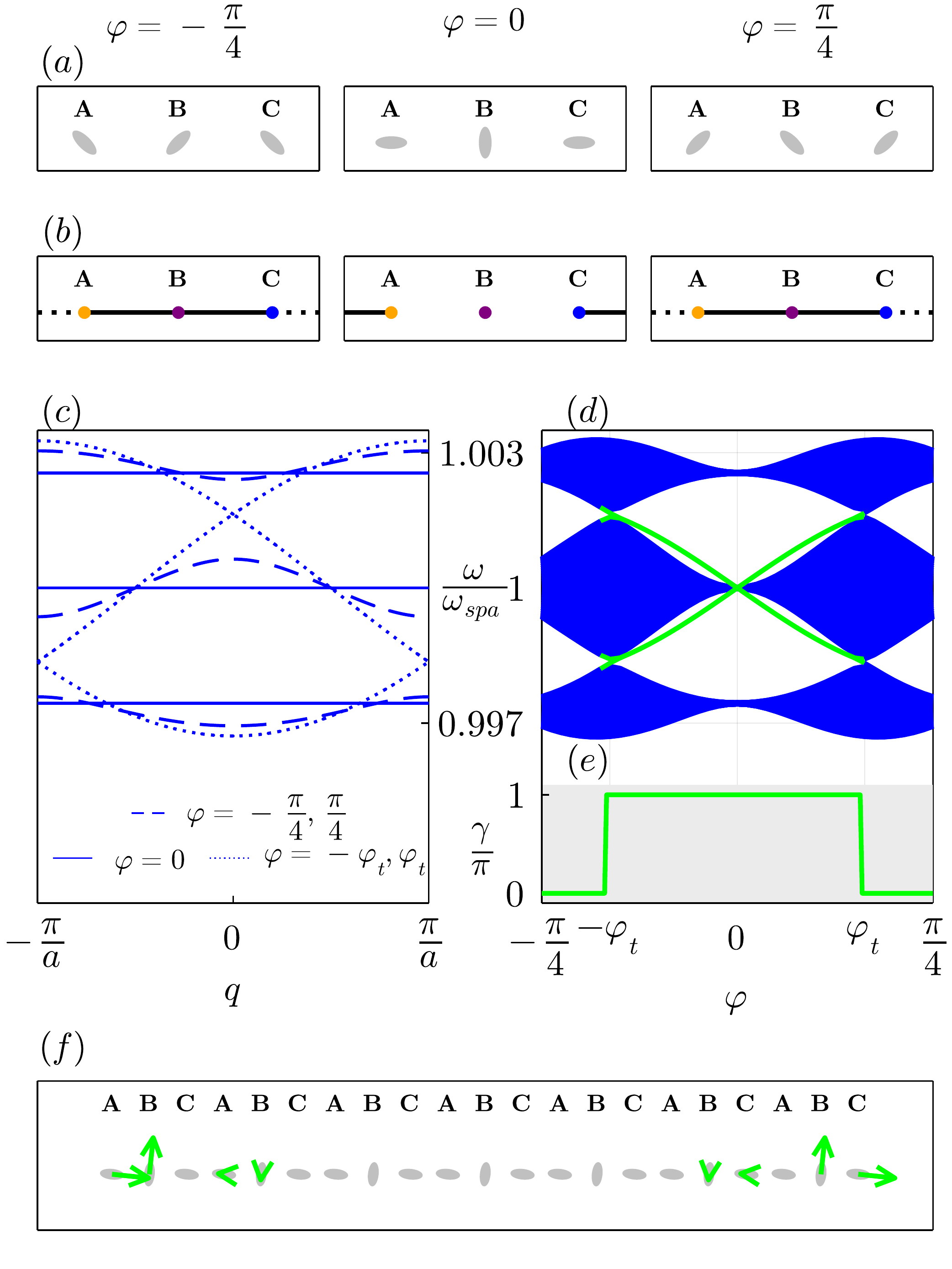}
\end{center}
\caption{\textbf{Plasmonic analogue of the mirror-symmetric SSH3 model}: periodic chain of 3 prolate silver nanospheroids per unit cell, with long spheroidal axes forming angles with the chain direction $\varphi_A=\varphi_B-\frac{\pi}{2} = \varphi_C = \varphi$. The dimensions of the nanoparticles are $a = 12.5 \text{ nm}$ and $b=c = 0.4a = 5 $ nm.
(a) Unit cells for $\varphi=-\frac{\pi}{4},$ $\varphi=0,$ $\varphi=\frac{\pi}{4}$. For any $\varphi$ the unit cell is inversion symmetric. (b) Equivalent tight binding unit cells. Solid and dashed black lines strong and weak hoppings. (c) Plasmonic bands of the periodic system. Solid lines represent the bands for $\varphi= 0$, dashed ones represent the bands for $\varphi = -\frac{\pi}{4}, \frac{\pi}{4}$ and dotted lines are $\varphi = -\varphi_t, \varphi_t$, with $\varphi_t \sim 0.16\pi$, where the lower and upper gaps close and topological transitions occur. (d) Plasmonic spectrum of a finite chain of 99 nanoparticles (33 unit cells). Bulk states are represented by blue lines, while green curves represent the two pairs of three-way-chiral edge states, that appear between $-\varphi_t$ and $\varphi_t$. (e) Zak phase of lower/upper gaps, which matches  with the existence of edge states in panel (d). (f) Edge state for $\varphi = -\frac{\pi}{30}$. Due to the three-way chirality the edge state is localized at the two sublattices closer to each edge.}
\label{fig:SSH3bands}
\end{figure} 
However, if we enlarge the unit cell, we can open a gap in an array of equidistant particles. Let us consider a 3-particle unit cell with $R=\frac{d}{3}$. Then the Green dyadic's matrix is:
\begin{align} 
\mathcal{G}(q) =  \begin{pmatrix} 0 & G_{\textbf{u}_A, \textbf{u}_B}& G_{\textbf{u}_A, \textbf{u}_C}  e^{-iqd} \\G_{\textbf{u}_A, \textbf{u}_B}& 0 & G_{\textbf{u}_B, \textbf{u}_C} \\ G_{\textbf{u}_A, \textbf{u}_C}e^{iqd} & G_{\textbf{u}_B, \textbf{u}_C}&  0 \end{pmatrix},
\end{align}
which is equivalent to the Bloch Hamiltonian of the SSH3 model (EQ. \ref{eq:BlochSSH3}).
The condition for the unit cell to be mirror-symmetric is $\varphi_{A} = -\varphi_C$ and $\varphi_B = 0, \frac{\pi}{2}$. The condition for the unit cell to be inversion-symmetric is, however, less restrictive. As a single nanospheroid is inversion-symmetric, the only condition is that particles $A$ and $C$ are inversion-symmetric with respect to each other, that is $\varphi_A = \varphi_C$.

The condition for $\mathcal{G}(q)$ to be equivalent to the mirror-symmetric SSH3 is  $|G_{\textbf{u}_A,\textbf{u}_B}| = |G_{\textbf{u}_2,\textbf{u}_3}|$. This is satisfied by any two pair of angles $\varphi_A, \varphi_B$ and $\varphi_B, \varphi_C$ that lay on the same or opposite contour line in Fig. \ref{fig:nanoparticles}(d). Explicitly, this occurs for $\varphi_B = \arctan\left(\frac{2\left(\cos{\varphi_A}\pm\cos{\varphi}_C\right)}{\left(\sin{\varphi_A} \pm \sin{\varphi_C}\right)}\right)$. 
This includes mirror symmetric and inversion symmetric previous conditions. However, it goes beyond them. For example, if we fix $\varphi_A=0, \varphi_C=\frac{\pi}{4}$, then the Green dyadic is accidentally mirror symmetric for $\varphi_B \sim 0.43\pi$ and $\varphi_B \sim 0.78\pi$. 

This accidental symmetry stems from the fact that we are ignoring the orientation of the nanoparticles in the equations, so it's a symmetry of the strength of the interaction between particles. Due to the anisotropy between longitudinal and transversal modes, these symmetries happen for apparently random values of the orientations. However, this accidental symmetry is enough to quantize the Zak phase, as in the true mirror symmetric and inversion symmetric cases.

In FIG. \ref{fig:SSH3bands} we show the topological transition in this plasmonic analogue of the mirror-symmetric SSH3 model with $a=12.5$nm, $d=15a$ and $R=\frac{d}{3}=5a$. By orienting the nanoparticles at $\varphi_A=\varphi_B+\frac{\pi}{2} =\varphi_C=\varphi$, upper and lower gaps close when $|G_{u_C,u_A}|=|G_{u_A,u_B}|=|G_{u_B,u_C}|$, that is, for $\varphi=\pm\varphi_t\sim\pm 0.16\pi$. For $|\varphi|> \varphi_t$ and $|G_{u_C,u_A}|<|G_{u_A,u_B}|$,  the system is in the trivial phase, while for $|\varphi|< \varphi_t$ and $|G_{u_C,u_A}|>|G_{u_A,u_B}|$  there are non-zero edge states in the upper/lower gaps. We can see the gaps closing and reopening for the bands of the periodic chain in panel (c) and for the finite chain (d). After the closing, double degenerate topological edge states (red solid lines in panel (d)) arise at the edges of the chain, localizing in odd sublattices at left edges and even sublattices at right edges due to sublattice symmetry. The appearance of these edge states matches the steps in the Zak phase, shown in panel (e). We also show one of the edge states of the lower gap for $\varphi_D=-\varphi_B = \frac{\pi}{4}$ in FIG.~\ref{fig:SSH3bands}(f), which as we see inherits the symmetry of the three-way chirality, localizing in two of the sublattices.

Interestingly, the asymmetry of transversal and longitudinal Green dyadic's functions can be exploited not just to open a gap but also to engineer accidental spatial symmetries or to suppress interaction between particles. This system is therefore more flexible than the linear and zigzag chains of nanospheres, allowing to play with symmetries, which yield a response of the edge states tunable by the external electric field, to be studied in last section.

\subsubsection{4 particles/unit cell}
\begin{figure}[h]
\begin{center}
\includegraphics[width=0.5\textwidth]{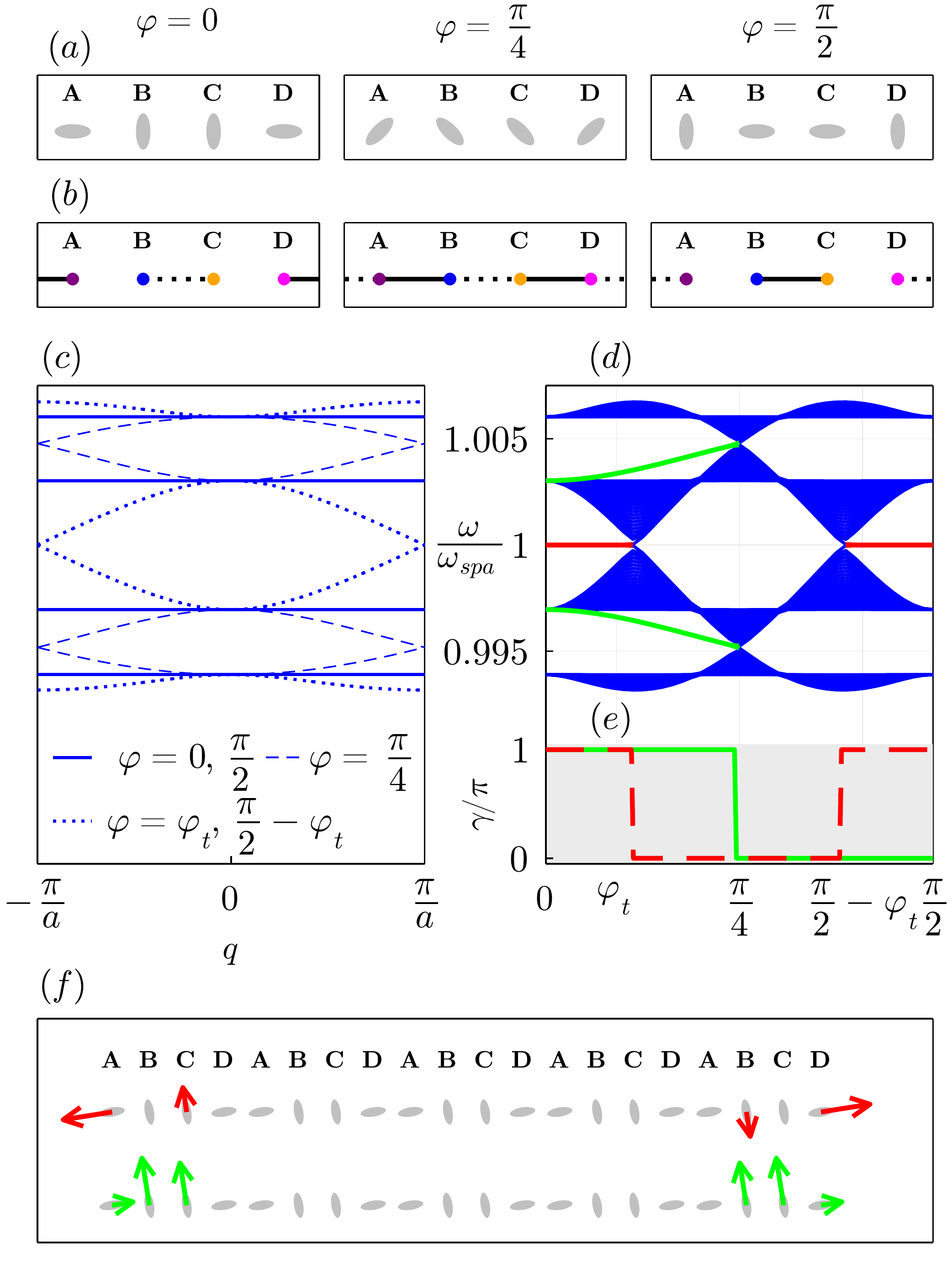}
\end{center}
\caption{\textbf{Plasmonic analogue of the SSH4 model}: periodic chain of 4 prolate silver nanospheroids per unit cell, with major axes forming angles with the chain direction $\varphi_A=\varphi_B-\frac{\pi}{2} =\varphi_C-\frac{\pi}{2} =\varphi_D = \varphi$. The dimensions of the nanoparticles are $a = 12.5 \text{ nm}$ and $b=c = 0.4a = 5 $ nm.
(a) Unit cells for $\varphi=0, \varphi=\frac{\pi}{4}$ and $\varphi=\frac{\pi}{2}$. For all values of $\varphi$, the unit cell is mirror-symmetric, but for $\varphi=0,\frac{\pi}{2}$ it is also inversion-symmetric. (b) Equivalent tight binding unit cells. Solid and dashed black lines represent weak and strong hoppings. (c) Plasmonic bands of the periodic system. Solid curves represent the bands for $\varphi = 0, -\frac{\pi}{2}$, whereas dash ones represent the bands for $\varphi_B = 0$, at the gap closing. (d) Plasmonic spectrum of a finite chain of 100 nanoparticles (25 unit cells). Bulk states are represented by blue lines, while red lines represent the pair of edge states, that appear after the gap closing at $\varphi_B = 0$. (e) Zak phases of central gap (solid red line) and lower/upper gaps (dashed green line), which match with the number of edge states in panel (d). The Zak phase of the central gap is quantized by chiral and mirror/inversion symmetries, while the lower and upper gap Zak phases are quantized just by mirror/inversion symmetries. (f) Edge state for $\varphi_D = -\varphi_B = \frac{\pi}{4}$. Due to sublattice symmetry left edge chiral states (red arrows) localize at odd sublattices ($A$ and $C$) while right edge state localize at even sublattices ($B$ and $D$), while four-way chiral edge states (green arrows) are localized in the closest three sublattices to the edge.}
\label{fig:SSH4bands}
\end{figure} 
Next, we consider a larger linear unit cell of 4 nanoparticles separated by a distance $R=\frac{d}{4}$. For this system, generally, $\mathcal{G}(q)$ in the base ($A,C,B,D$) is:
\begin{equation} 
\begin{pmatrix} 0  & 0 &G_{\textbf{u}_A, \textbf{u}_B}&  G_{\textbf{u}_A, \textbf{u}_D}e^{-iqd} \\ 0 & 0 &  G_{\textbf{u}_B, \textbf{u}_C} &G_{\textbf{u}_C, \textbf{u}_D} \\ 
G_{\textbf{u}_A, \textbf{u}_B} & G_{\textbf{u}_B, \textbf{u}_C} & 0 & 0  \\G_{\textbf{u}_A, \textbf{u}_D}e^{iqd}& G_{\textbf{u}_C, \textbf{u}_D} & 0 & 0\end{pmatrix}. \end{equation}

This is equivalent to the Hamiltonian of the SSH4 model (EQ.~\ref{eq:BlochSSH4}) with hoppings given by EQ.~\ref{eq:eqGth1d}. As we see, this matrix is block-antidiagonal, i.e. sublattice symmetric. This is because there are two sublattices (odd and even sites) with only inter-sublattice connections.

The conditions for this system to be geometrically mirror (inversion) symmetric are $\varphi_1=\mp\varphi_4$ and $\varphi_2=\mp\varphi_3$. 
However, the only condition for $\mathcal{G}(q)$ to be effectively mirror-symmetric is  $|G_{\textbf{u}_A, \textbf{u}_B}|  = |G_{\textbf{u}_C, \textbf{u}_D}|$. This is satisfied by any two pairs of $\varphi_A, \varphi_B$ and $\varphi_C, \varphi_D$ that lie in the same or opposite contour line. True or hidden mirror/inversion symmetries quantize the Zak phase of all gaps.
\begin{figure}[h]
\begin{center}
\includegraphics[width=0.5\textwidth]{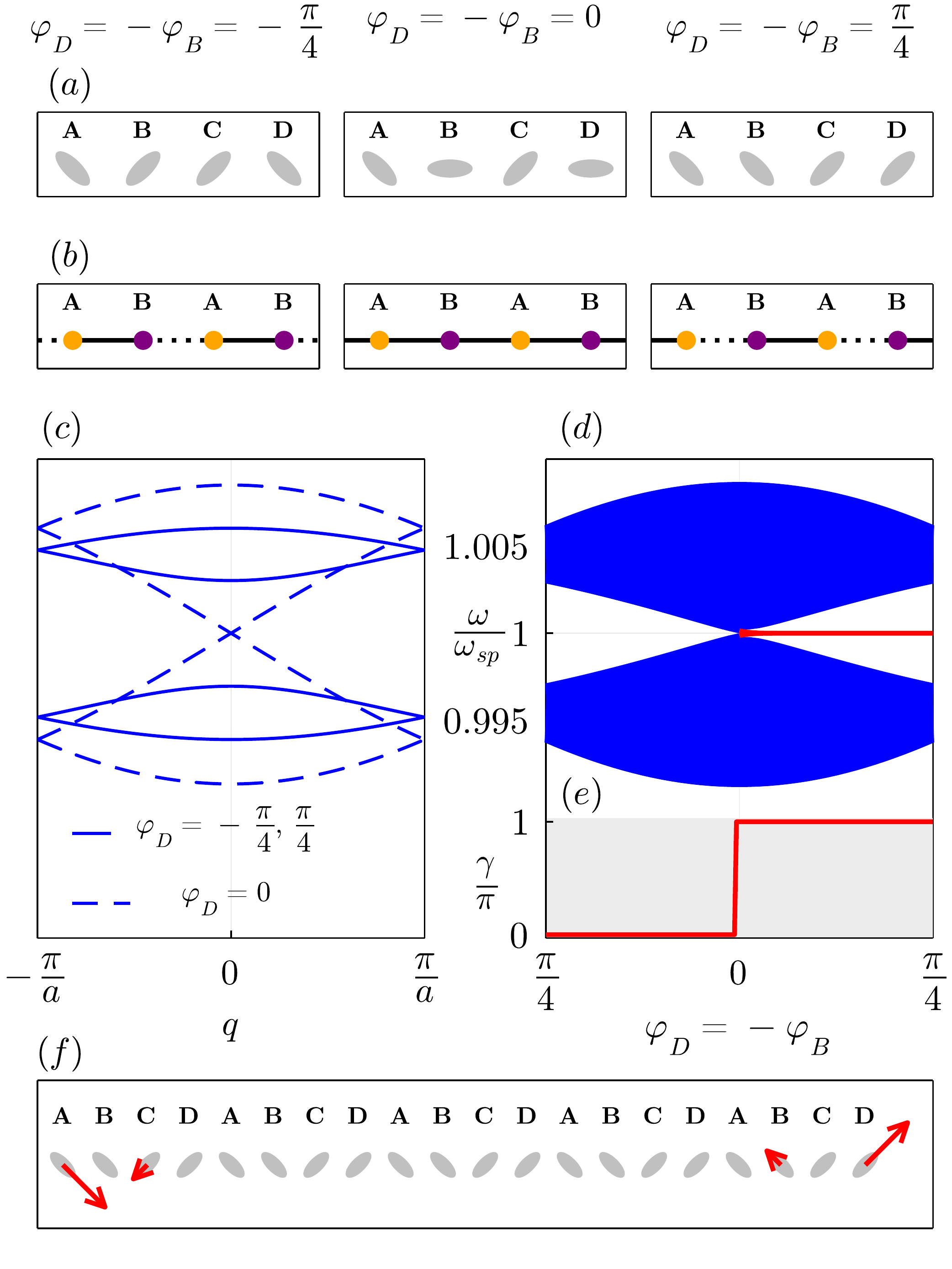}
\end{center}
\caption{\textbf{Plasmonic analogue of the SSH model}: periodic chain of 4 prolate silver nanospheroids per unit cell, with long spheroidal axes forming angles with the chain direction $\varphi_A=\varphi_C = \frac{\pi}{4}$ and $\varphi_B = -\varphi_D$ let free. The dimensions of the nanoparticles are $a = 12.5 \text{ nm}$ and $b=c = 0.4a = 5 $ nm.
(a) Unit cells for  $\varphi_D=-\pi/4,0,\pi/4$. (b) Equivalent tight binding unit cells; the effective unit cells are diatomic, as in the SSH model. Solid and dashed black lines represent strong and weak bonds. (c) Plasmonic bands of the periodic system. Solid curves represent the bands for $\varphi_B = -\frac{\pi}{4}, -\frac{\pi}{4}$, while dashed ones represent the bands for $\varphi_B = 0$, at the gap closing. (d) Plasmonic spectrum of a finite chain of 100 nanoparticles (25 unit cells). Bulk states are represented by blue lines, while red lines represent the pair of edge states, that appear after the gap closing at $\varphi_B = 0$. (e) Zak phase of central gap, which matches  the number of edge states in panel (d). (f) Edge state for $\varphi_D = -\varphi_B = \frac{\pi}{4}$. Due to sublattice symmetry, the left edge state is localized at odd sublattices ($A$ and $C$) while the right edge state is localized at even sublattices ($B$ and $D$).}
\label{fig:SSHbands}
\end{figure} 
\begin{figure*}[h]
\begin{center}
\includegraphics[width=0.45\textwidth]{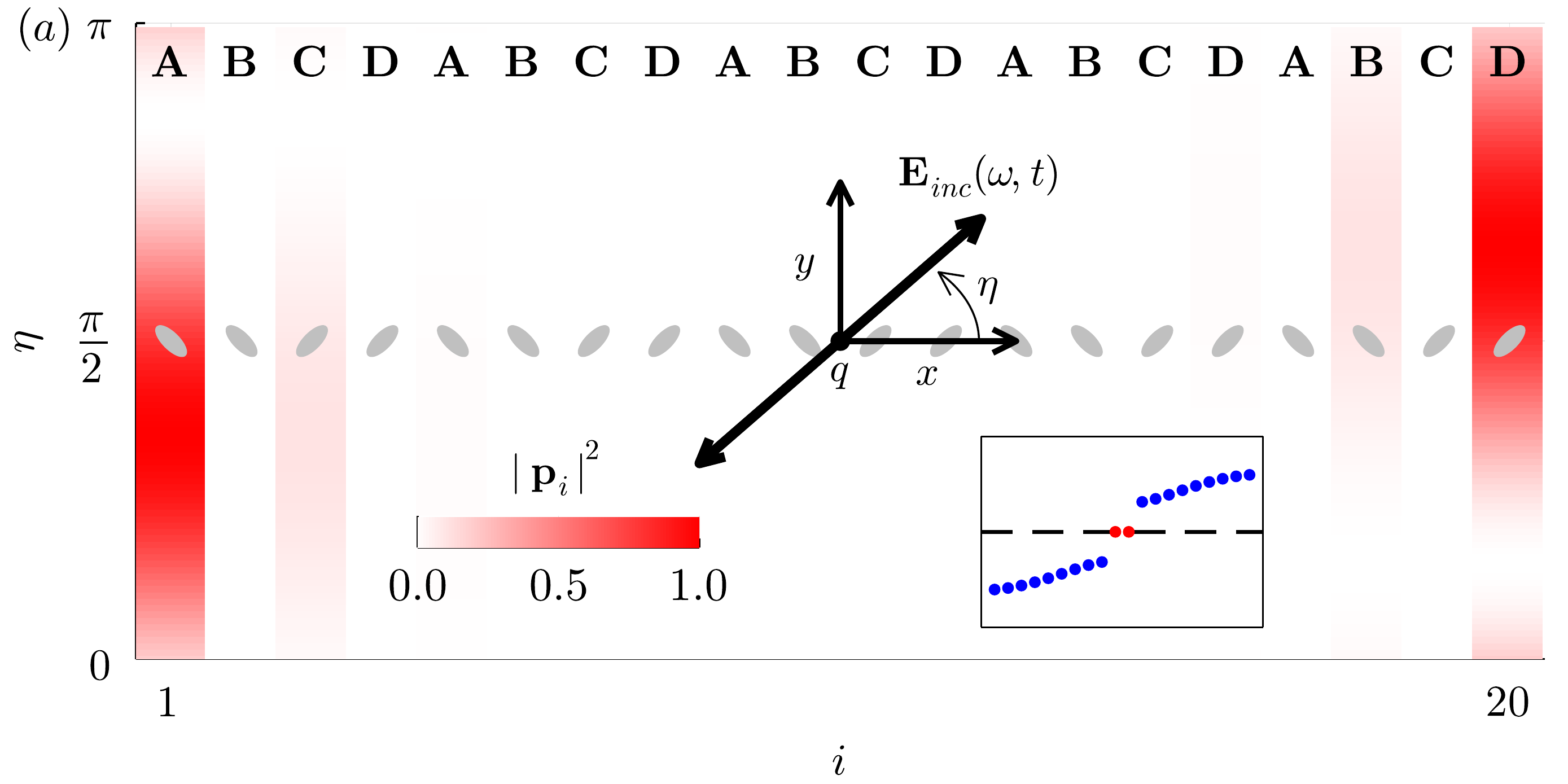}
\includegraphics[width=0.45\textwidth]{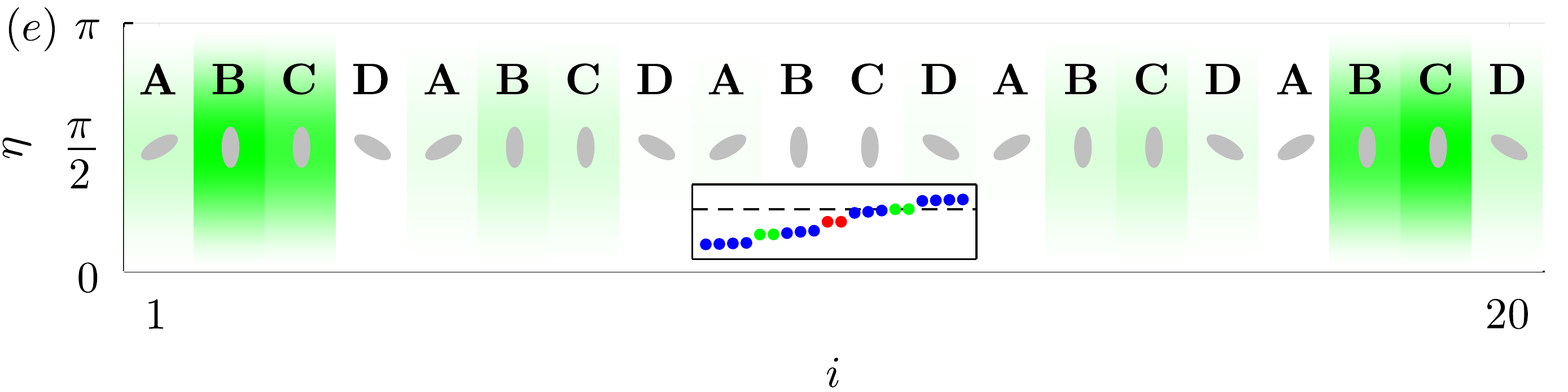}
\includegraphics[width=0.45\textwidth]{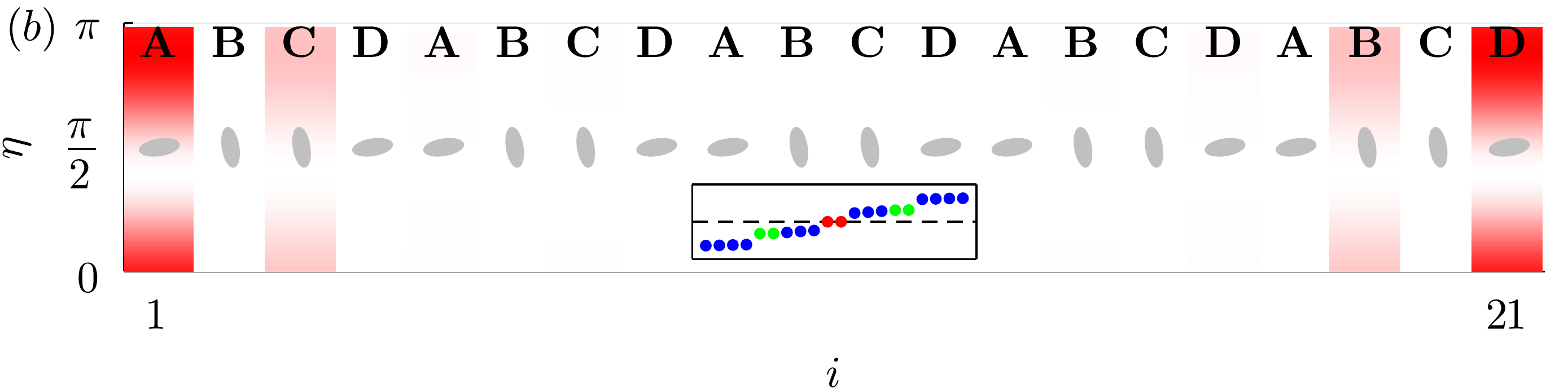}
\includegraphics[width=0.45\textwidth]{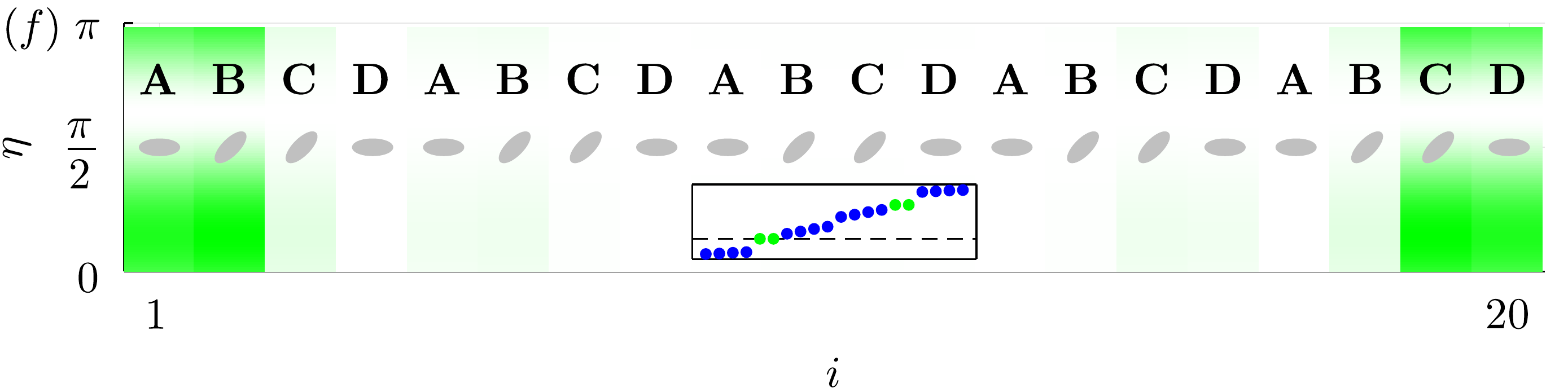}
\includegraphics[width=0.45\textwidth]{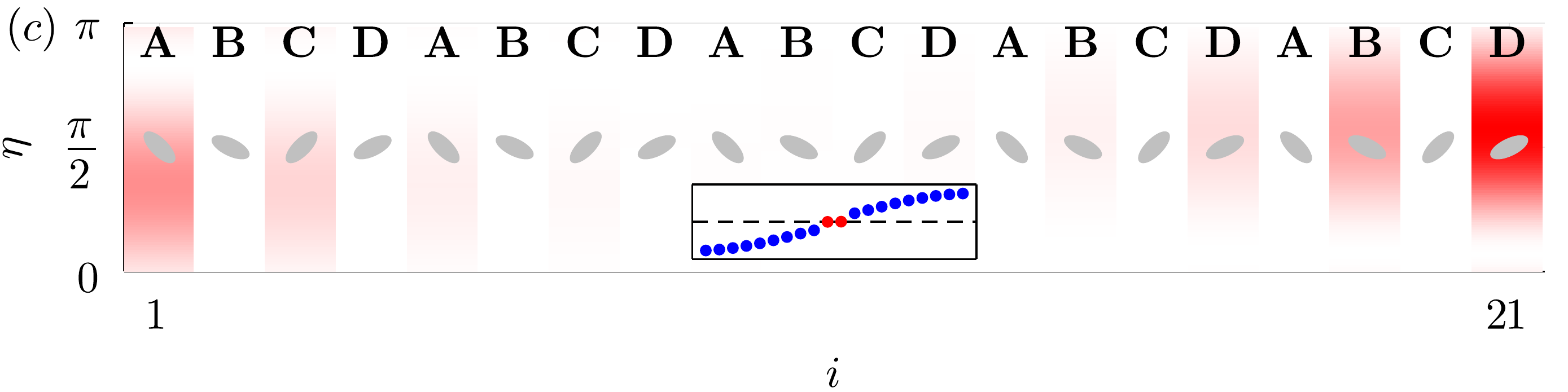}
\includegraphics[width=0.45\textwidth]{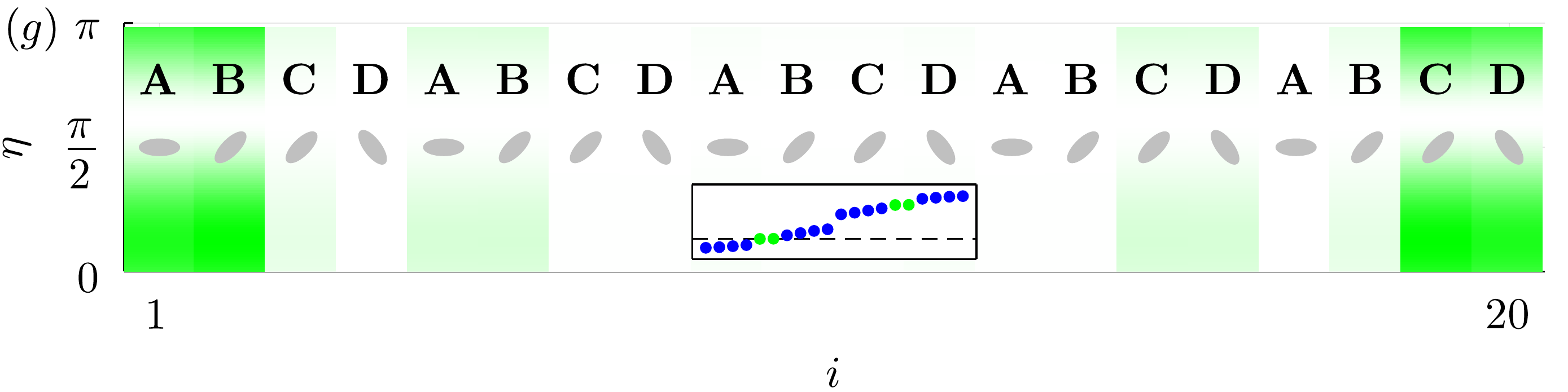}
\includegraphics[width=0.45\textwidth]{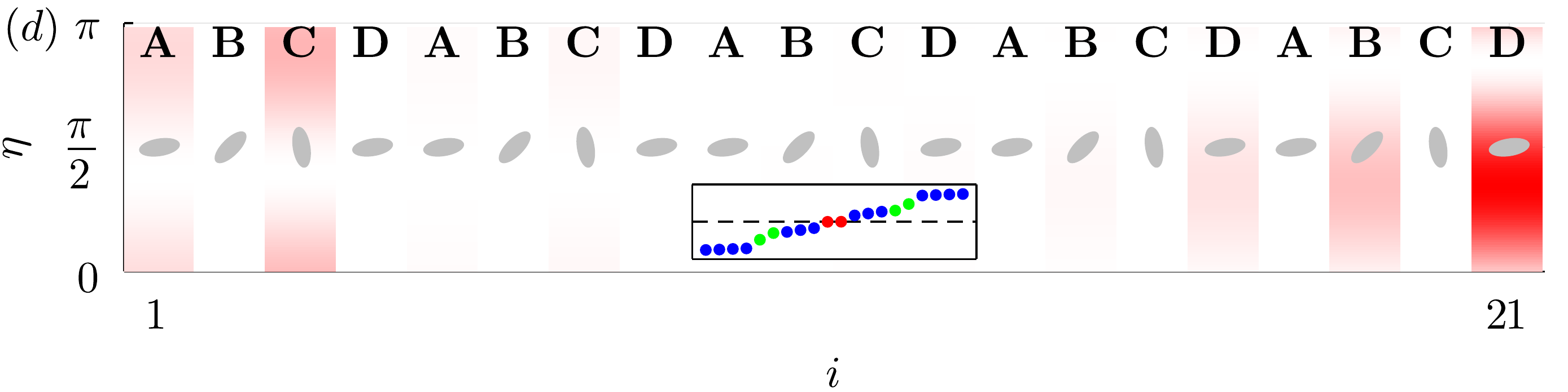}
\includegraphics[width=0.45\textwidth]{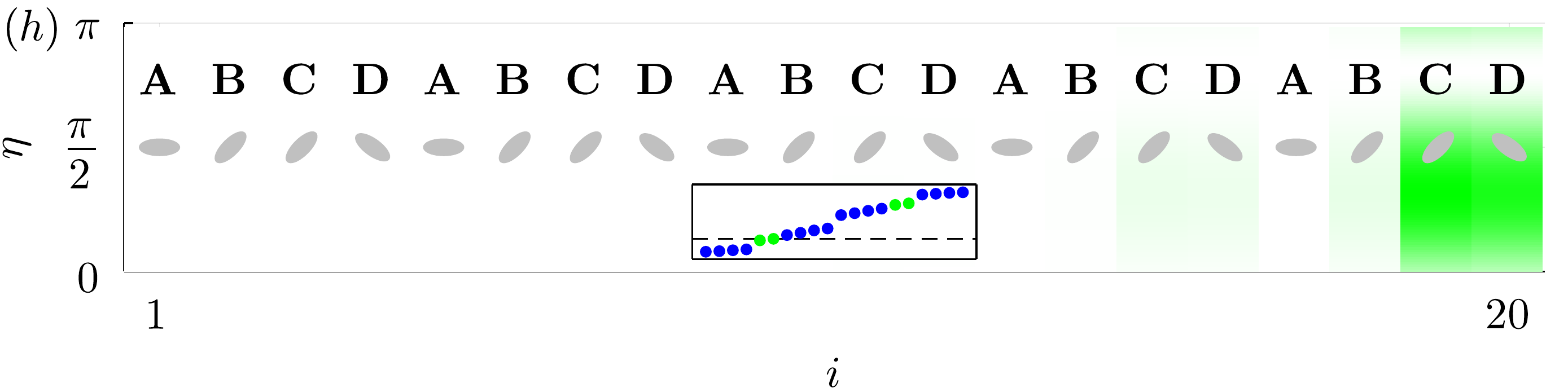}
\end{center}
\caption{\textbf{Excitation of edge states of different plasmonic chains by an incoming linearly polarized electric field at normal incidence, depending on its angle of polarization $\eta$}. We plot the square of the module of dipolar moments at each site $\textbf{p}_i$, normalized by its maximum value for all polarizations and sites. We use red and green gradients for chiral (central gap) in panels (a-d) and generalized 4-way chiral (lower gap) edge states in panels (e-h). As insets we plot the spectra of the chains, where blue, red and green dots represent bulk states, central gap edge states, and lower/upper gap edge states. (a) Mirror symmetric unit cell (see Fig 5),$\varphi_n=\pi/4[-1,-1,1,1]$. The chain breaks inversion symmetry, while incident electric field breaks mirror symmetry (except for $\eta=0$ and $\eta=\pi$), allowing to switch off left or right edge states separately for $\eta \sim \frac{\pi}{7}$ and $\eta \sim \frac{6\pi}{7}$. Then, when applying a circular incident electric field, edge states bounce back and forth between the edges. (b) Inversion symmetric unit cell. Electric field doesn't break inversion symmetry, but allows to switch on and off both edge states simultaneously. (c) Accidentally mirror symmetric unit cell. As both true mirror and inversion symmetries are broken, both the amplitude and phase of the edge states differ. (d) Non-spatial-symmetric unit cell. As left-right edge states are still degenerate due to chiral symmetry, we have a response for both edges similar to case (c). (e) Mirror symmetric unit cell.  (f,g) Accidentally mirror symmetric unit cells. (h) Non-spatial-symmetric unit cell. Due to the absence of symmetries, left and right edge states are not degenerate and can be excited separately.}
\label{fig:externalfield}
\end{figure*} 
In FIG. \ref{fig:SSH4bands} we show a plasmonic analogue of the SSH4 model with $a=12.5$nm, $d=15a$ and $R=\frac{d}{4}=3.75a$. By starting from the unit cell with $\varphi_A=\varphi_B-\pi/2=\varphi_C-\pi/2 =\varphi_D = 0$ (left unit cell in panel (a)) and rotating all nanoparticles an angle $\varphi$,  two different topological transitions are crossed. As we see in panels (c) and (d), the first one occurs at $\varphi=\varphi_t \sim 0.12\pi$ ($\varphi=\frac{\pi}{2}-\varphi_t$), where $|G_{\textbf{u}_A, \textbf{u}_B}G_{\textbf{u}_C, \textbf{u}_D}| = |G_{\textbf{u}_B, \textbf{u}_C}G_{\textbf{u}_D, \textbf{u}_A}|$ and the central gap closes. After reopening, the edge states (red solid line in panel (d)) disappear (appear). For $\varphi=0$, $G_{\textbf{u}_D, \textbf{u}_A}= G_{\textbf{u}_B, \textbf{u}_C}$ and $G_{\textbf{u}_A, \textbf{u}_B} = G_{\textbf{u}_C, \textbf{u}_D}$, so upper and lower gaps close and its edge states (green solid line in panel (d)) disappear after the closing, when $G_{u_D,u_A} < G_{u_B,u_C}$. In panel (e) we show the Zak phase for all the gaps, which represent the existence of edge states in each gap. 
In panel (f) we plot the dipoles for the edge states in the central (red arrows) and lower  (green arrows) gaps. As we see, the former respects the sublattice symmetry and is localized only in odd or even sublattices, while the latter has only  zero weight in one of the sublattices.

Interestingly, due to accidental symmetries, in this system we can also recover the topology of the SSH model. When $G_{\textbf{u}_D, \textbf{u}_A}= G_{\textbf{u}_B, \textbf{u}_C}$ and $G_{\textbf{u}_A, \textbf{u}_B} = G_{\textbf{u}_C, \textbf{u}_D}$ (or equivalently, $t=v$ and $u=w$ in FIG.~\ref{fig:tbchains}(c)), upper and lower gaps close and we have an analogue of the SSH model. Even when the period of the real unit cell is $4R$, the effective tight binding unit cell has a period of $2R$.

In FIG. \ref{fig:SSHbands} we see a possible realization of this analogue of the SSH for $a=12.5$nm, $d=15a$ and $R=\frac{d}{4}=3.75a$. By fixing the direction of the nanoparticles in sites $A$ and $C$ at $\varphi_A=-\varphi_C =-\frac{\pi}{4}$ and rotating $\varphi_B = -\varphi_D$ from $-\frac{\pi}{4}$ (left unit cell in FIG.~\ref{fig:SSHbands}(a)) to $\frac{\pi}{4}$ (right unit cell), a topological transition occurs at $\varphi_B=-\varphi_D =-\frac{\pi}{4}=0$ (middle unit cell), where the gap closes due to all $G_{\textbf{u}_i, \textbf{u}_j}$ being equivalent, as in the SSH model for $v=w$ (middle unit cell in panel (b)). In this system there's also a transition from an inversion symmetric unit cell $(\varphi_B=-\frac{\pi}{4})$ to an accidentally mirror symmetric one $(-\frac{\pi}{4}<\varphi_B<\frac{\pi}{4})$ to a mirror symmetric unit cell  $(\varphi_B=\frac{\pi}{4})$, so the equivalent tight binding unit cells (panel (b)) always remain mirror symmetric, as in the SSH. We can see the gap closing and reopening for the bands of the periodic chain. (panel (c)) and for the finite chain (panel (d)). We also plot the Zak phase (panel (e)), that compared to that of panel (d), we see it represents the existence of edge states (red line). After the closing, double degenerate topological edge states arise at the edges of the chain, localizing in odd sublattices at the left edge and in even sublattices at the right edge due to sublattice symmetry. We show one of the edge states for $\varphi_D=-\varphi_B = \frac{\pi}{4}$ in panel (f). 

In the next section we will study how these edge states are excited by an incident electric field, depending on its polarization.

\section{Switching edge states by incoming electric field}
\label{section:Excitation}
A difference between electronics and plasmonics is that plasmons are not fermions, so the bands are not naturally half-filled. In photonics, we need an incident field that overlaps spatially with the eigensolutions of the array.
Once the incident field is fixed, by inverting EQ.~\ref{eq:dipoles}, the dipoles of the chain are given by:
\begin{align}
\textbf{P} =  \left(\mathcal{G}(\omega) - \frac{1}{\alpha(\omega)}I\right)^{-1}\textbf{E}_{inc}(\omega,t) ,
\label{eq:DipolesField}
\end{align}
$\textbf{E}_{inc}$ being a $3 \times N$ vector which contains the field $E_{inc}(t)$ evaluated at the position of each nanoparticle.
For a chain of nanospheroids, the equation reduces to: 
\begin{align}
\textbf{P}_a =  \left(\mathcal{G}_a(\omega) - \frac{1}{\alpha_a(\omega)}I\right)^{-1}\textbf{E}^a_{inc}(\omega,t),
\end{align}
where $\textbf{E}_{inc}^a$ is a vector of the projections of the electric field in the directions of the major axes of the nanoparticles. Such projections are key in order to excite or not protected states.
When all the particles in the array are spherical or oriented in the same direction, the electric fields affect almost equally all the nanoparticles. However, when nanoparticles are oriented in different directions, the external electric field can couple to real and hidden spatial symmetries.

Let's analyze what happens when we excite the edge states of the plasmonic chains. In FIG.~\ref{fig:externalfield} we plot the dipolar response to a linearly polarized electric field at normal incidence, depending on its polarization:
\begin{equation} 
\textbf{E}_{inc}(\omega,t) \propto  \begin{pmatrix} \cos\eta \\ \sin\eta \\ 0 \end{pmatrix},
\end{equation} 
where $\eta$ is the angle of polarization of the electric field with respect to the $x$ axis. In order to the field to resonate with the nanoparticles and with the edge state mode, due to the losses of the nanoparticles, we need the incoming electric field to have the frequency of the edge state and a finite lifetime, that is, a pulse. 

In FIG.~\ref{fig:externalfield}(a-d) we analize the edge states of the central gap, which are protected by chiral symmetry. In panels (e-h) we excite the edge states in the lower gap of the SSH4 chain, which are 4-way-chiral-symmetric. For both types we consider chains with mirror, inversion, accidental mirror, and no spatial symmetries to see how this affects the optical response. 

In FIG.~\ref{fig:externalfield}(a) we see the response of the SSH nanospheroid chain with mirror symmetry and $-\varphi_A=-\varphi_B=\varphi_C=\varphi_D = \frac{\pi}{4}$ to a linearly polarized electric field at the frequency of the surface plasmon $\omega_{spa}$, depending on its polarization. Since all the nanoparticles are oriented at diagonals when the field is polarized in $x$ or $y$ directions $(\eta=0,\frac{\pi}{2})$, all the particles are equally perturbed so mirror symmetry holds, and left and right edge states are identical. 

However, when we apply an electric field oriented at $\eta \neq 0, \frac{\pi}{2}$, the interaction with the external field depends on the nanoparticle. The sublattice symmetry is still preserved, as we see in Fig \ref{fig:externalfield} (a). However, the external field breaks the mirror symmetry, allowing to have a different response at left and right edges. For $\eta \simeq  \pi/7$ ($\eta \simeq  \frac{6\pi}{7}$), the dipolar response is localized only at the left (right) edge. Then, by changing the polarization of the field we can select left, right or both edge states with the same or different weight.  

Now if we apply a circularly polarized electric field at normal incidence, this is:
\begin{equation}
\textbf{E}_{inc}(\omega,t) \propto \begin{pmatrix} \cos(\eta(t)) \\ \sin(\eta(t)) \\ 0 \end{pmatrix}.
\end{equation}
The nanospheroids convert the circular polarization of the incoming field to linear polarization.  Then, the oscillations in left and right edges are not in phase, so the edge states "bounce back and forth" between left and right edges. Over a period T, the response of the chain loops two times over the $\eta$ axis in FIG.~\ref{fig:externalfield}(a).

However, when the chain is inversion-symmetric, for example the one in panel (c) ($\varphi_n = \pi/8 + [0,\pi/2,\pi/2,0]$), the electric field preserves this symmetry, so the response in both edges is the same. We can switch on/off both edge states simultaneously. If the electric field is circularly polarized, then the oscillations in both edges are in phase. 

When the chain is accidentally mirror-symmetric (panel (c)), or has no spatial symmetries, the field couples more intensely to one of the edges. If we apply a circularly polarized electric field, the oscillations in the edges would not be just dephased as the bouncing states in the mirror symmetry chain, but they would differ also in amplitude.

For the edge states in lower (or upper) gaps, however, we find a different scenario. In a mirror symmetric SSH4 chain (panel (e), $\varphi_n = [\pi/4, \pi/2, \pi/2, \pi/4]$), the external field doesn't appear to break the symmetry between edges. This may be due to the coexistence of generalized chiral symmetry and spatial symmetries. This means that we cannot select right or left edges. If the field is circularly polarized, then the oscillations in both edges are in phase.
The same occurs for an inversion-symmetric unit cell (panel (f), $\varphi_n = [0, \pi/4, \pi/4, \pi/4,0]$) and in the accidentally symmetric case (panel (g), $\varphi_n = [0, \pi/4, \pi/4, 0.71\pi])$). 

Finally, if the SSH4 unit cell has no spatial symmetries (panel (h), $\varphi_n = [0, \pi/4, \pi/4, 0.8\pi]$), left and right lower/upper gap edge states are no longer degenerate, so we can excite them separately at different frequencies. However, these edge states don't have any kind of topological protection and can be pushed out of the gap by disorder and hybridize with bulk states.

As we see,  by using elongated nanoparticles we have gained control in edge states, making possible to switch them off, select left, right, both, or bouncing edge states. This wasn't feasible in the nanosphere chain, as all the particles were identical and have isotropic response for all the polarizations of the electric field. 

\section{Conclusions}
In previous years, there have been several proposals to mimic topological electronic systems in photonics. Periodic arrays of metallic nanoparticles are an interesting platform to study topology in nanophotonics due to their plasmonic resonances in the visible range and their tunability. Here we have proposed means to open a topological gap not by rearranging the particles in an array as in crystalline topological electronic systems, but by orientating elongated particles. By adding this degree of freedom, we can mimic topological chains as the SSH model or its greater unit cell extensions in an equidistant array. The spatial polarization modulation allows also to switch on/off or select right, left or bouncing edges states, by changing the polarization of the incoming electric field, as in the zigzag plasmonic chain. However, this system is even more flexible, making possible also to suppress interaction between nanoparticles or to engineer accidental spatial symmetries. In this paper we proved that orientation of nanoparticles can open a topological gap in an otherwise gapless system (which  could also be exploited in gapped $1D$ and $2D$ arrays), allowing to filter modes or play with symmetries and polarizations. This opens a path towards exploiting features of nanoparticles for topology without a counterpart in condensed matter systems. 
 
\bibliography{references}
\clearpage
\appendix
\onecolumngrid
\section{Polarizability of metallic nanospheres and nanospheroids}
Generally, the electric polarizability $\overleftrightarrow{\alpha}(\omega)$ acts like a tensor:
\begin{align}
\overleftrightarrow{\alpha}(\omega) = \begin{pmatrix} \alpha_{xx}  & \alpha_{xy}  & \alpha_{xz} \\ \alpha_{xy} & \alpha_{yy} & \alpha_{yz} \\ \alpha_{xz} & \alpha_{yz} & \alpha_{zz} \end{pmatrix} .
\end{align}
For a nanosphere, due to the spherical symmetry, the polarizability is proportional to the identity matrix, $\overleftrightarrow{\alpha}(\omega)= \alpha(\omega)I$. When we are working in the limit $a\gg\lambda$ we can take the quasi-static approximation, assuming only the first Mie coefficient contributes to the polarizability, so $\alpha(\omega)$ is: 
\begin{align}
\alpha(\omega) = 4\pi a^3 \epsilon_0 \frac{\epsilon(\omega) - \epsilon_\mathrm{B}}{\epsilon(\omega) + 2\epsilon_\mathrm{B}}. 
\label{eq:alpha_QS2}
\end{align}
The permittivity of the medium, $\epsilon(\omega)$ can be approximated by a Drude-Lorentz model \cite{Rodrigo2008}: 
\begin{equation}
\epsilon(\omega) = \epsilon_r - \sum_j \frac{\omega^2_{\mathrm{P},j}}{\omega(\omega + i\gamma_j)} - \sum_j \frac{\Delta\epsilon_j\Omega^2_{j}}{\omega^2 - \Omega^2_j +  i\omega\Gamma_j},
\end{equation}
where $\epsilon_r$ is the static dielectric constant, $\omega_{\mathrm{P},j}$ are plasma frequencies, $\gamma_j$ and $\Gamma_j$ are the damping constants, $\Omega_j$ are resonant frequencies, and $\Delta\epsilon_j$ are related to the oscillator strengths. In our work we use the parameters for silver: $\epsilon_r =4.6, \; \omega_{\mathrm{P},0} = 9.0, \; \gamma_0 = 0.07,\; \Gamma_0 =1.2, \; \Omega_0 = 4.9, \; \Delta \epsilon_0 = 1.10$ \cite{Rodrigo2008}.

The extinction cross section is the sum of absorption and scattering cross sections and for a single nanoparticle as given by \cite{novotny12}:
\begin{equation}
\sigma_{\mathrm{ext}} = \sigma_{\mathrm{abs}} + \sigma_{\mathrm{sca}}= \frac{k}{\epsilon_0}\textnormal{Im}{(\alpha(\omega)}) + \frac{k^4}{6\pi\epsilon_0^2}|\alpha(\omega)|^2. 
\end{equation}

Now we consider prolate spheroids, that is, ellipsoids with major axis $a$ and minor axes $b=c$. The polarizability doesn't behave like a scalar anymore, but depends on the polarization of the incoming electric field. The polarizabilities for the main axes, $\alpha_l(\omega)$ with $l \in [a,b,c]$, are \cite{Moroz2009}:
\begin{equation}
\alpha_{l}(\omega) = V \frac{\epsilon(\omega)-\epsilon_b}{\epsilon_b + L_l(\epsilon(\omega)-\epsilon_b)} ,
\end{equation}
$V$ being  the volume of the spheroid, $V = \frac{4}{3}\pi a c^2$, and $L_l$ are geometric factors given by:
\begin{align}
L_a = \frac{e^2}{1-e^2}\left(\frac{1}{2\sqrt{1-e^2}}\ln\left(\frac{1+\sqrt{1-e^2}}{1-\sqrt{{1-e^2}}}\right) -1\right), \nonumber \\
L_b = L_c = \frac{1-L_a}{2},
\end{align}
where $e = c/a$ is the eccentricity of the spheroid, ranging from $e=0$ (needle) to $e=1$ (sphere).

Let us now consider an array of nanoparticles, where the major axis of the spheroid $n$ is in the direction $\textbf{u}_n = (\sin{\theta_n}\cos{\varphi_n},\sin{\theta_n}\sin{\varphi_n}, \cos{\theta_n})$, where $\theta_n$ and $\varphi_n$ are the angles formed by the spheroidal major axis with respect to the $z$ and $x$ axis. When $q\ll1$, in the vicinity of the major axis resonance $\omega_{spa}$,  $\alpha_c(\omega \simeq \omega_{spa})\simeq 0$,  we can approximate the polarizability tensor as:
\begin{equation}
\overleftrightarrow{\alpha_n}(\omega) \simeq \alpha_a(\omega) \begin{pmatrix} \sin^2 \theta_n \cos^2 \varphi_n & \sin^2 \theta_n \sin \varphi_n \cos \varphi_n & \sin \theta_n \cos \theta_n \cos \varphi_n \\ \sin^2 \theta_n\sin \varphi_n \cos \varphi_n & \sin^2 \theta_n \sin^2 \varphi_n & \sin \theta_n \cos \theta_n \sin \varphi_n \\ \sin \theta_n \cos \theta_n \cos \varphi_n & \sin \theta_n \cos \theta_n \sin \varphi_n & \cos^2 \theta_n \end{pmatrix} ,
\end{equation}
which projects any vector $\textbf{v}$ in the direction $\textbf{v}$, i.e. $\overleftrightarrow{\alpha}(\omega)\textbf{v} = \alpha_a(\omega)(\textbf{v}\cdot \textbf{u}_n)\cdot \textbf{u}_n$. 

As the direction of the dipoles is fixed, we can project $\textbf{p}_n$ in the direction $u_n$, so we get the scalar coupled-dipole equations in EQ.~\ref{eq:eqG2}.

\label{section:AppxA}

\section{Green dyadic's function projection}
\label{section:AppxB}

As we show in appendix \ref{section:AppxA}, near $\omega_{spa}$ the polarizabilities of the nanospheroids project the dipoles in the directions of the major axes. This turns in scalar coupled-dipole equations (see EQ.~\ref{eq:eqG2}). Explicitly, the Green's dyadic function projection for two nanospheroids with major axes oriented in $\textbf{u}_n = (\sin{\theta_n}\cos{\varphi_n},\sin{\theta_n}\sin{\varphi_n}, \cos{\theta_n})$ and $\textbf{u}_m, (\sin{\theta_m}\cos{\varphi_m},\sin{\theta_m}\sin{\varphi_m}, \cos{\theta_m})$ directions, $G_{\textbf{u}_m, \textbf{u}_n}$,  is: 
\begin{equation}
\begin{split}
G_{\textbf{u}_m, \textbf{u}_n} = 
G_{xx} \cos{\varphi_m}\cos{\varphi_n}\sin{\theta_m}\sin{\theta_n} + G_{xy} \sin{\theta_m}\sin{\theta_n}\sin{(\varphi_m+\varphi_n)} + G_{xz}\left(\cos{\theta_m}\cos{\varphi_n}\sin{\theta_n}  + cos{\theta_n}\cos{\varphi_m} \sin{\theta_m} \right) + \nonumber \\ + G_{yy} \sin{\varphi_m}\sin{\varphi_n}\sin{\theta_m}\sin{\theta_n}  +G_{yz}\left(\cos{\theta_m}\sin{\varphi_n}\sin{\theta_n}  + cos{\theta_n}\sin{\varphi_m} \sin{\theta_m} \right) + \nonumber \\ +  G_{zz} \cos{\theta_m}\cos{\theta_n},
\label{eq:Gdproj}
\end{split}
\end{equation}

where $\mu,\nu = x,y,z$ are the polarizations and $G_{\mu\nu}$ are the elements of the Green dyadic $\GG(\textbf{r}_{m},\textbf{r}_{n},\omega)$. 

For a linear array of nanoparticles along the $x$ direction, EQ.~\ref{eq:Gdproj} reduces to:
\begin{eqnarray} 
G_{\textbf{u}_m, \textbf{u}_n} = \frac{(2\cos{\varphi_m}\cos{\varphi_n} - \sin{\varphi_m}\sin{\varphi_n})\sin{\theta_m}\sin{\theta_n}  - \cos{\theta_m}\cos{\theta_n}}{4\pi k^2 R^3},
 \end{eqnarray}
which for $\theta_n = \theta_m =  \frac{\pi}{2}$ derives in EQ.~\ref{eq:eqGth1d}. In this paper, we restricted the orientations of the dipoles to the $xy$ or $xz$ planes, but by orienting them in the space, we could add another extra degree of freedom. 

\end{document}